\documentclass[prd,showpacs]{revtex4}
\usepackage{epsfig}
\usepackage{graphicx}

\newcommand{\gsim}{\lower.7ex\hbox{$\;\stackrel{\textstyle>}{\sim}\;$}}
\newcommand{\lsim}{\lower.7ex\hbox{$\;\stackrel{\textstyle<}{\sim}\;$}}

\begin{document}

\title{Lightest Higgs Boson and Relic Neutralino in the MSSM with CP Violation}

\author{Jae Sik Lee$^{\dagger}$ and Stefano Scopel$^*$}
\affiliation{
$^{\dagger}$  Center of Theoretical Physics, School of
Physics, Seoul National University, Seoul, 151--747, Korea\\
$^*$ Korea Institute for Advanced Study, Seoul
130-722, Korea}


\begin{abstract}
We discuss the lower bound to the lightest Higgs boson $H_1$ in the
minimal supersymmetric extension of the standard model (MSSM) with explicit
CP violation, and the phenomenology of the lightest relic neutralino
in the same scenario. In particular, adopting the CPX benchmark
scenario, we find that the combination
of experimental constraints coming from LEP, Thallium Electric Dipole
Moment (EDM) measurements, quorkonium decays, and $B_s\rightarrow \mu\mu$ decay
favours a region of the
parameter space where the mass of $H_1$ is in the range $7~{\rm GeV}\lsim
M_{H_1}\lsim 10$ GeV, while $3 \lsim \tan\beta\lsim 5$.
Assuming a departure from the usual GUT relation among gaugino masses
($|M_1| \ll |M_2|$), we find that through resonant annihilation to
$H_1$ a neutralino
as light as 2.9 GeV can be a viable dark matter candidate in this
scenario.  
We call this the CPX light neutralino scenario, and discuss its
phenomenology showing that indirect Dark Matter searches
are compatible with the present experimental constraints, as long as
$m_\chi\lsim M_{H_1}/2$. On the other hand, part of the range
$m_\chi\gsim M_{H_1}/2$ which is allowed by cosmology is excluded by
antiproton fluxes.

\end{abstract}

\maketitle

\baselineskip 1.5em

\section{Introduction}

Supersymmetry (SUSY) is considered one of the most natural extensions
of the Standard Model (SM), providing elegant solutions to puzzles as
diverse as the SM hierarchy problem, the high--energy unification of
the gauge coupling constants and the existence of the Dark Matter (DM)
in the Universe. In particular, in R--parity conserving SUSY scenarios
the lightest neutralino turns out to be an ideal thermal DM
candidate~\cite{susy_review}, providing in a natural way the correct
amount of Cold Dark Matter (CDM) that is needed to drive structure
formation, and which is necessary to explain the latest data on the
energy budget of the Universe from WMAP~\cite{wmap3}.

Unfortunately, our ignorance about the details of the SUSY breaking
mechanism implies that phenomenological analysis on the SUSY DM
depend in general on a huge parameter space with more than 100
independent soft masses and couplings. This parameter space is usually
drastically resized by making use of as many simplifying assumptions
as possible. For instance, parameters related to flavour--mixing are
supposed to be strongly suppressed to match the experimental
constraints, and are often neglected. On the other hand, in
theoretically motivated setups as in specific SUSY--breaking
scenarios, like in the minimal Supergravity (SUGRA) ~\cite{SUGRA}, the
number of free parameters is strongly reduced, improving
predictability. However, some other assumptions are simply suggested by
simplicity, such as taking soft--breaking parameters real.

In particular, many recent analysis have addressed this latter
aspect~\cite{APLB,CPHIGGS,HeinCP,CPNSH,
CPsuperH_ELP,INhiggs,Gondolo_CP,Nihei,Belanger}, since it has been realized
that CP violating phases in the soft terms can considerably enrich the
phenomenology without violating existing constraints. This is also
theoretically motivated by the fact that the smallness of neutrino
masses implied by observation possibly calls for some exotic source of
CP violation additional to Yukawa couplings, in order to explain
Baryogenesis.

In the CP--conserving SUGRA scenario, the so--called
``stau--coannihilation'', ``Higgs--funnel'' and ``focus point''
benchmark solutions are well known examples of a situation where,
thanks to a combination of different experimental constraints, quite
simple and well defined phenomenological pictures emerge, at the price
of a certain amount of tuning ~\cite{SUGRA}.  In this article we point
out that a similar situation occurs in an effective Minimal
Supersymmetric extension of the Standard Model (MSSM) with all soft
parameters fixed at the electro--weak scale, when CP violation and
departure from unification of gaugino masses are considered.

In particular, we wish to address here the issues of the lower bound
for the mass of the lightest Higgs boson $H_1$ and that of the lightest
possible mass for the relic neutralino $\chi$, when standard
assumptions are made for the origin and evolution of its relic
density.  In fact, by combining present experimental constraints, a
very simple picture (albeit tuned) emerges, where the mass of the
lightest Higgs boson $H_1$ is found to be in the range $7\lsim M_{H_1}
\lsim$ 7.5 GeV, with the ratio of the two vacuum expectation values
almost fixed, $\tan\beta\simeq$ 3. This range can be relaxed to:
$7\lsim M_{H_1}\lsim$ 10 GeV and $3\lsim \tan\beta\lsim$ 5, with quite
mild assumptions on the Thallium EDM. In this scenario, resonant
annihilations of neutralinos with mass $m_{\chi}\simeq M_{H_1}/2$
through $H_1$ exchange in the $s$ channel can drive their thermal
relic abundance within the limits coming from observation, for values
of $m_\chi$ significantly below those allowed in CP--conserving
scenarios ~\cite{mchi_lowerbound} (to our knowledge, the issue of
resonant $\chi$ annihilation in the context of CP violation was first
raised at the qualitative level in ~\cite{Gondolo_CP}).

In the following we will analyse in detail the implications for direct
and indirect DM searches of these light neutralinos, which we will
refer to as the CPX light neutralino scenario, concluding that it is
indeed viable scenario, with prospects of detection in future
experiments.

The plan of the paper is as follows. In Section II the CPX scenario
in the MSSM with CP violation is introduced. In Section III we discuss
various experimental bounds. Section IV is devoted to the discussion
of the cosmological relic density of CPX light relic neutralinos, and
Section V to their phenomenology in DM searches. Our conclusions are
contained in Section VI.

\section{MSSM with explicit CP violation: the CPX scenario}

The tree level Higgs potential of the MSSM is invariant under CP
transformations. However, CP can be explicitly broken at the loop
level.
In the presence of sizable CP phases in the relevant soft SUSY breaking
terms, a significant mixing between the scalar and pseudo--scalar
neutral Higgs bosons can be generated
~\cite{APLB,CPHIGGS,HeinCP}.  As a consequence of this CP--violating mixing,
the three neutral MSSM Higgs mass eigenstates, labeled in order of
increasing mass as $M_{H_1}\le M_{H_2} \le M_{H_3}$, have no longer
definite CP parities, but become mixtures of CP-even and CP-odd
states.  In this scenario, all masses are usually calculated as a
function of the charged Higgs boson mass $M_{H^\pm}$, instead of the
pseudoscalar Higgs mass $M_A$, which is no longer a physical parameter. Much
work has been devoted to studying the phenomenological features of this
radiative Higgs-sector CP violation in the framework of the MSSM
~\cite{CPNSH,CPsuperH_ELP}.

Due to the large Yukawa couplings, the CP-violating mixing among the
neutral Higgs bosons is dominated by the contribution of
third-generation squarks and is proportional to the combination:
\begin{equation}
\frac{3}{16\pi^2}\frac{\Im{\rm
m}(A_f\,\mu)}{m_{\tilde{f}_2}^2-m_{\tilde{f}_1}^2}\label{eq:ratio},
\end{equation}
\noindent with $f=t,b$. Here $\mu$ is the Higgs--mixing parameter in
the superpotential and $A_f$ denotes the trilinear soft coupling.  In
particular, the amount of CP violation is enhanced when the product of
$|A_{b,t}|$ and $|\mu|$ is larger than the difference of the sfermion
masses squared.  At the two--loop level, also the gluino mass
parameter $M_3$ becomes relevant through threshold corrections to the
top- and bottom-quark Yukawa couplings. This contribution depends on
the combination ${\rm Arg}(M_3\,\mu)$ and can be important especially
when $\tan\beta\equiv v_2/v_1$ is large, where $v_2$ and $v_1$ are the
vacuum expectation values of the neutral components of the two Higgs
doublets that give masses to up--type and down--type quarks,
respectively. More CP phases become relevant by including subdominant
radiative corrections from other sectors ~\cite{INhiggs}.

In presence of CP violation, the mixing among neutral Higgs bosons is
described by a 3$\times$3 real orthogonal matrix $O$, instead of a
2$\times$2 one.  The matrix $O$ relates the electro--weak states to the
mass eigenstates as:
\begin{equation}
(\phi_1\,,\phi_2\,,a)^{T}=O\,(H_1\,,H_2\,,H_3)^T\,.
\label{eq:omix}
\end{equation}
\noindent 
We note that the elements $O_{\phi_1 i}$ and $O_{\phi_2 i}$ are 
the CP-even components of the $i$-th Higgs
boson, while $O_{a i}$ is the corresponding CP-odd component.

The Higgs-boson couplings to the SM and SUSY particles could be modified
significantly due to the CP violating mixing.
Among them, one of the most important ones may be the Higgs-boson coupling to a pair
of vector bosons, $g_{H_iVV}$, which is responsible for the production of Higgs
bosons at $e^+e^-$ colliders:
\begin{equation}
{\cal L}_{HVV}=gM_W\left(W_\mu^+W^{-\mu}+\frac{1}{2c_W^2}Z_\mu Z^\mu\right)
\sum_{i=1}^3 g_{H_iVV}H_i\,,
\label{eq:hvvinterx}
\end{equation}
where
\begin{equation}
g_{H_iVV}=c_\beta \, O_{\phi_1i}+s_\beta \, O_{\phi_2 i}\,,
\end{equation}
\noindent when normalized to the SM value. Here we have used the following abbreviations:
$s_\beta\equiv\sin\beta$, $c_\beta\equiv\cos\beta$.
$t_\beta=\tan\beta$, etc.
We note that the two vector bosons $W$ and $Z$ couple only to the
CP-even components $O_{\phi_{1,2} i}$ of the $i$-th Higgs mass
eigenstate, and the relevant couplings may be strongly
suppressed when the $i$-th Higgs boson is mostly CP-odd, $O^2_{ai}\sim
1 \gg O_{\phi_1i}^2\,,O_{\phi_2i}^2$.

The so called CPX scenario has been defined as a showcase benchmark
point for studying CP-violating Higgs-mixing phenomena ~\cite{CPX}. Its
parameters are all defined at the electro--weak scale, and are chosen
in order to enhance the combination in Eq.(\ref{eq:ratio}).
In this scenario,
SUSY soft parameters are fixed as follows:
\begin{eqnarray}
&& \hspace{-2cm}
M_{\tilde{Q}_3} = M_{\tilde{U}_3} = M_{\tilde{D}_3} =
M_{\tilde{L}_3} = M_{\tilde{E}_3} = M_{\rm SUSY}\,,
\nonumber \\
&& \hspace{-2cm}
|\mu|=4\,M_{\rm SUSY}\,, \ \
|A_{t,b,\tau}|=2\,M_{\rm SUSY} \,, \ \
|M_3|=1 ~~{\rm TeV}\,,
\label{eq:mssm-intro-CPXdef}
\end{eqnarray}
\noindent where, with a usual notation, $Q$, $L$, $U$, $D$ and $E$
indicate chiral supermultiplets corresponding to left-- and
right--handed quarks and leptons. In this scenario $\tan\beta$,
$M_{H^\pm}$, and $M_{\rm SUSY}$ are free parameters.
As far as CP phases are concerned, we adopt, without loss of generality, the
convention ${\rm Arg}(\mu)=0$, while we assume a common phase for all
the $A_f$ terms, $\Phi_A\equiv {\rm Arg}(A_t)={\rm Arg}(A_b)={\rm
Arg}(A_\tau)$.  As a consequence of this, we end--up with two free
physical phases: $\Phi_A$ and $\Phi_3={\rm Arg}(M_3)$.

In addition to the parameters fixed by the CPX scenario,
we need to fix the gaugino masses $M_{1,2}$ for our study. 
We take them as free parameters
independently of $M_3$ since, for them, we chose to relax the usual
relations at the electro-weak scale:
$M_i/M_j=g_i^2/g_j^2$ with $g_{i,j}$=gauge coupling constants, 
which originate from the
assumption of gaugino--mass unification at the GUT scale.  
The neutralino $\chi$ is defined as usual as the lowest-mass linear
superposition of  $B$-ino $\tilde{B}$, $W$-ino $\tilde{W}^{(3)}$, and of the
two Higgsino states $\tilde{H}^0_1$, $\tilde{H}^0_2$:

\begin{equation}
\chi\equiv a_1 \tilde{B}+a_2\tilde{W}^{(3)}+a_3 \tilde{H}^0_1+a_4 \tilde{H}^0_2.
\end{equation}

\noindent

In Ref.~\cite{light_neutralinos} it was proved that in a
CP--conserving effective MSSM with $|M_1| << |M_2|$ light neutralinos
of a mass as low as 7 GeV are allowed. Indeed, for $|M_1| << |M_2|$
the LEP constraints do not apply, and the lower bound on the
neutralino mass is set by the cosmological bound. As shown in
\cite{light_neutralinos}, these neutralinos turn out to be mainly
$B$-inos, $a_1\simeq 1$ and $m_{\chi}\simeq |M_1|$, with a small
Higgsino component given by:
\begin{equation}
\frac{|a_3|}{|a_1|}\simeq \sin\theta_W\sin\beta \frac{M_Z}{\mu},
\label{eq:a3_a1}
\end{equation}
 \noindent where $\theta_W$ is the Weinberg angle and $M_Z$ is the
Z--boson mass. In the following we will assume vanishing phases for
$M_1$ and $M_2$, and we will fix for definiteness $M_2$=200 GeV (the
phenomenology we are interested in is not sensitive to these
parameters in a significant way). On the other hand, we will vary
$M_1$, which is directly correlated to the lightest neutralino mass
$m_{\chi}$.

In this work, we rely on {\tt CPsuperH} ~\cite{CPsuperH} for the computation of
mass spectra and couplings in the MSSM Higgs sector.

\section{Experimental constraints on the CPX scenario}

\subsection{LEP2 searches}
The most relevant feature of the CPX scenario for our analysis is
that the lightest Higgs boson $H_1$ can be very light, $M_{H_1}\lsim$ 10 GeV,
with the other two neutral Higgs bosons significantly heavier, $M_{H_{2,3}}\gsim$ 100 GeV,
when $\Phi_A \sim 90^{\rm o}$ and $M_{H^\pm} \sim 130$ GeV
for moderate values of $3 \lsim \tan\beta \lsim 10$.
In this case, the lightest Higgs boson is mostly CP odd and its
production at LEP is highly suppressed since $|g_{H_1 VV}|\ll 1$ 
though it is kinematically accessible.  On the other hand, the second--lightest 
Higgs $H_2$ can be produced together with a $Z$ boson since its 
complementary coupling $g_{H_2 VV}$ is sizeable. But its mass is close to the 
kinematical limit $\sim 110$ GeV and, moreover,
it dominantly decays into two $H_1$'s. Depending on $M_{H_1}$, the lightest
Higgs boson decays into two $b$ quarks or two $\tau$ leptons.
This leads to a dominant production and decay mode containing 6 jets in the final
state, a topology which was covered by LEP2 with a very low efficiency.
The similar situation occurs for $H_1$--$H_2$ pair production.
Therefore, in the presence of CP-violating mixings, a
very light Higgs bosons with $M_{H_1} \lsim$ 10 GeV could easily
escape detection at LEP2.  

For the CPX scenario, taking $\Phi_A=\Phi_3=90^\circ$ and $M_{\rm
SUSY}=0.5$ TeV, the combined searches of the four LEP collaborations
at $\sqrt{s}= 91 - 209$ GeV reported the following two
allowed regions~\cite{LEP_HIGGS}:
\begin{eqnarray}
    &{\bf R1}:&\;\;\; M_{H_1} \lsim 10 \;\; {\rm GeV}\;\;\;\; 
      {\rm for}\;\; 3 \lsim \tan\beta \lsim 10,\nonumber\\
    &{\bf R2}:&\;\;\; 30\;\; {\rm GeV} \lsim M_{H_1} \lsim 50\;\; {\rm
    GeV}\;\;\;\; {\rm for} \;\; 3  \lsim \tan\beta \lsim 10.
\label{eq:lep}
\end{eqnarray}
\noindent
These regions will not be fully covered even at the LHC for $\tan\beta
\lsim 7 \;({\bf R1})$ and $\lsim 5 \;({\bf R2})$ ~\cite{M.Schumacher}.
In our analysis we will focus on region ${\bf R1}$.

We observe that in the scenario analysed by the LEP collaborations one
has $|\mu|$=2 TeV. For this large value of $|\mu|$, the neutralino is
a very pure $B$-ino configuration, with a Higgsino contamination
$a_3\simeq$0.02 (see Eq. \ref{eq:a3_a1}). As will be shown in Section
\ref{section:relicdensity}, this has important consequences for the
phenomenology of relic neutralinos, in particular suppressing their
annihilation cross section, and restricting the possibility of having
a relic abundance in the allowed range only to the case of resonant
annihilation. So, the exploration of different possibilities with
lower values of $|\mu|$ could in principle be very relevant for relic
neutralinos.  However, this would require a re--analysis of LEP data
which is beyond the scope of this paper.

\begin{figure}[t]
\vspace{0.0cm}
\centerline{\epsfig{figure=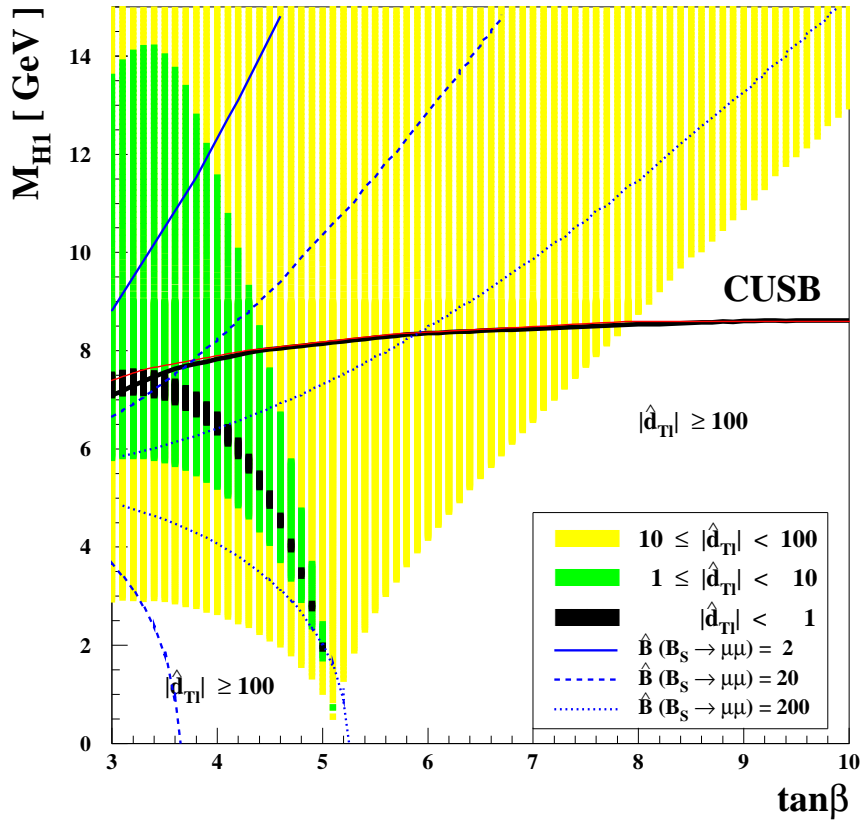,height=14cm,width=14cm}}
\vspace{-0.5cm}
\caption{{\it The Thallium EDM $\hat{d}_{\rm Tl} \equiv d_{\rm Tl}
\times 10^{24}$~$e\,cm$ in the CPX scenario with $M_{\rm SUSY}=0.5$ TeV
in the region $M_{H_1} \lsim 15$ GeV and $3 < \tan\beta < 10$.  The different shaded
regions correspond to different ranges of $|\hat{d}_{\rm Tl}|$, as
shown: specifically, the narrow region consistent with the current thallium
EDM constraint, $|\hat{d}_{\rm Tl}|<1$, is denoted by black squares.
In the blank unshaded region we have $|\hat{d}_{\rm Tl}|>100$. The region
below the thick solid line is excluded by data on $\Upsilon(1S)$ decay
\protect\cite{upsilon_visible}. For comparison, the thin line shows 
an estimation of the same boundary obtained using the tree-level coupling
taking $O_{a1}$=1, i.e. $|g^P_{H_1\bar{b}b}| = \tan\beta$. Also shown are
the three contour lines of 
the rescaled $\hat{B}(B_s\rightarrow \mu\mu)\equiv
(B_s\rightarrow \mu\mu)  \times 10^7$: $\hat{B}(B_s\rightarrow \mu\mu)=2$ (solid),
20 (dotted), and 200 (dashed).}}
\label{fig:dtl}
\end{figure}

\subsection{Electric Dipole Moments}
CP phases in the MSSM are significantly constrained by the EDM
measurements. In particular, the EDM of the Thallium atom provides
currently the most stringent constraint on the MSSM scenario of our
interest.  The atomic EDM of $^{205}$Tl gets its main contributions
from two terms ~\cite{KL,PR}:
\begin{equation}
d_{\rm Tl}\,[e\,cm]\ =\ -585\cdot d_e\,[e\,cm]\:
-\: 8.5\times 10^{-19}\,[e\,cm]\cdot (C_S\,{\rm TeV}^2)+ \cdots\,,
\end{equation}
where $d_e$ denotes the electron EDM and $C_S$ is the coefficient of
the CP-odd electron-nucleon interaction ${\cal
L}_{C_S}=C_S\,\bar{e}i\gamma_5 e\,\bar{N}N$.  The dots denote
sub-dominant contributions from 6-dimensional tensor and
higher-dimensional operators.  The above quantity is constrained
by the experimental $2-\sigma$ upper bound on the Thallium EDM, which
is~\cite{THEDMEXP}:
\begin{equation}
|d_{\rm Tl}|\ \lsim\ 1.3\times 10^{-24}\,[e\,cm]\; .
\end{equation}

The contributions of the first and second generation phases,
e.g.~$\Phi_{A_{e,\mu}}$, $\Phi_{A_{d,s}}$ etc., to EDMs can be
drastically reduced either by assuming these phases sufficiently
small, or if the first- and second-generation squarks and sleptons are
sufficiently heavy. However, even
when the contribution of the first and second generation phases to
EDM's is suppressed, there are still sizeable contributions to EDMs
from Higgs-mediated two loop diagrams ~\cite{CKP}. Their explicit forms
for $(d_e/e)^H$ and $C_S$ may be found in Ref.~\cite{Ellis:2005ik},
expressed in the conventions and notations of {\tt
CPsuperH}~\cite{CPsuperH}.

In Fig.~\ref{fig:dtl}, we show the rescaled Thallium EDM $\hat{d}_{\rm
Tl} \equiv d_{\rm Tl} \times 10^{24}$ in units of $e\,cm$ in the
$M_{H_1}$-$\tan\beta$ plane. Here, we consider only the
contributions from the Higgs-mediated two--loop diagrams.
Different ranges of $|\hat{d}_{\rm Tl}|$ are shown explicitly
by different shadings.  In the blank unshaded region we have obtained
$|\hat{d}_{\rm Tl}|>100$.
We find that a cancellation between the contributions from $(d_e/e)^H$
and $C_S$ occurs when $\tan\beta < 5$.  This cancellation is
responsible for the narrow region denoted by black squares with
$|\hat{d}_{\rm Tl}|<1$, where it is at the level of about about 5 \%.

Finally, we note that the Thallium EDM constraint can be evaded by
assuming cancellations between the two--loop contributions considered
here and other contributions, such as those from first-- and
second--generation sfermions discussed above.
In this way the allowed region shown in Fig.~\ref{fig:dtl} can be
enlarged.  The amount of cancellation can be directly read--off from
Fig.~\ref{fig:dtl}.  For instance, in the region $|\hat{d}_{\rm
Tl}|<10$ it would be less severe than 1 part in 10 (10 \%).

\subsection{Bottomonium decay}
\label{sec:quarkonium}

In the region {\bf R1}, see Eq.~(\ref{eq:lep}), 
the bottomonium decay
channel $\Upsilon(1S)\rightarrow \gamma H_1$ is kinematically
accessible ~\cite{higgs_hunter_guide}.
There are two experimental
limits on this process, depending on whether the $H_1$ decays to
visible particles ~\cite{upsilon_visible} or to invisible
ones~\cite{upsilon_invisible}. The second case is allowed when
2$m_{\chi}<M_{H_1}$, so that $H_1$ can decay to neutralinos which
escape detection. On the other hand, in the case 2$m_{\chi}>M_{H_1}$
also the three--body decay (i.e. with a non--monochromatic $\gamma$
spectrum) $\Upsilon(1S)\rightarrow \gamma \chi\chi$ has been
constrained ~\cite{upsilon_visible}. The branching ratio for the
two--body decay calculated in our scenario is related to its SM
counterpart by~\cite{higgs_hunter_guide}:

\begin{equation}
B(\Upsilon(1S)\rightarrow \gamma\, H_1)_{SUSY}=
B(\Upsilon(1S)\rightarrow \gamma\, H_1)_{SM}\times (g^P_{H_1\bar{b}b})^2\,,
\label{eq:upsilon}
\end{equation}

\noindent where $g^P_{H_1\bar{b}b}$ denotes the Higgs coupling to two
$b$ quarks given by $g^P_{H_1\bar{b}b}=-O_{a1}\tan\beta$ at the tree level. 
This implies that
the experimental upper bounds on this process can be directly
converted to a constraint in the plane $\tan\beta$--$M_{H_1}$.  The
result is shown in Fig.~\ref{fig:dtl}, where the thin (red) line corresponds
to the limit obtained by setting
$(g^P_{H_1\bar{b}b})^2=\tan^2\beta$. Finite--threshold corrections
induced by the gluino and chargino exchanges can modify the coupling
$g^P_{H_1\bar{b}b}$, although this effect is negligible at low values
of $\tan\beta$. Moreover, although for our choice of parameters $H_1$
is mostly pseudoscalar, $O_{a1}$ can be smaller than 1 up to about
20\%. The thick solid line in Fig.~\ref{fig:dtl} shows the bottomonium
constraint when the threshold corrections and $O_{a1}$ are fully included.

From Fig.~\ref{fig:dtl} one can see that, when the following
constraints are combined: (i) the LEP constraint; (ii) Thallium EDM; 
(iii) the limit from bottomonium decay,
the allowed parameter space is reduced to:
\begin{equation}
7 \;\;{\rm GeV} \lsim  M_{H_1}\lsim 7.5 \;\;{\rm GeV}
~~{\rm and}~~ \tan\beta\simeq 3.
\label{eq:tanb_mh1}
\end{equation}
\noindent This region may be enlarged to
\begin{equation}
7 \;\;{\rm GeV} \lsim  M_{H_1}\lsim 10 \;\;{\rm GeV}
~~{\rm and}~~ 3\lsim \tan\beta\lsim 5\,,
\label{eq:tanb_mh1_enlarged}
\end{equation}
\noindent if we assume 10 \%-level cancellation in the
Thallium EDM.

In light of the above discussion and for definiteness, from now on we
will fix $M_{H_1}$=7.5 GeV and $\tan\beta$=3 in our analysis. Taking
into account the CPX parameter choice of Eq.(\ref{eq:mssm-intro-CPXdef})
with  $\Phi_A=\Phi_3=90^\circ$ and $M_{\rm SUSY}=0.5$ TeV,
this implies, in particular: $M_{H^{\pm}}\simeq$ 147 GeV,
$M_{H_2}\simeq$ 108 GeV, $M_{H_3}\simeq$ 157 GeV.

\subsection{Other constraints}
\label{sec:other}

As will be discussed in the following sections, if the pseudoscalar
Higgs boson mass is in the range (\ref{eq:tanb_mh1}), a CPX light
neutralino with $m_{\chi}\lsim M_{H_1}/2$ can be a viable DM
candidate.  Due to their very pure $B$-ino composition, and to the
quite low value of $\tan\beta$, neutralinos in this mass range evade
constraints coming from accelerators. For instance, in the CPX light
neutralino mass range the present upper bound to the invisible width
of the $Z$--boson implies $|a_3^2-a_4^2|\lsim$ a few percent, a
constraint easily evaded in this case.

As far as Flavour Changing Neutral Currents (FCNC) are concerned, they
strongly depend on the assumptions about flavour violation in the
squark sector. For instance, assuming squarks diagonal in flavour, the
SUSY contribution to the $b\rightarrow s\gamma$ decay rate is
dominated by chargino--stop and $H^{\pm}$--W loops, which are strongly
suppressed in our case by the low value of $\tan\beta$ and by the fact
that there are no light masses to compensate this.

The situation is potentially different in the case of the decay
$B_s\rightarrow \mu\mu$, since its dominant SUSY contribution scales
as $\tan^6\beta\,|\mu|^2/M_{H_1}^4$ and may have a resonance
enhancement when $H_1$ is so light that $M_{H_1}\sim M_{B_s}$.
Neglecting the threshold corrections which are not so important in our
case, we estimate the branching ratio based on the approximated
expression~\cite{Ibrahim:2002fx}
\begin{equation}
B(B_s\rightarrow \mu\mu)\simeq
\frac{2\,\tau_{B_s}\,M_{B_s}^5\,f_{B_s}^2}{64\pi} \left|C
\right|^2(O_{\phi_1 1}^4+O_{a 1}^4)
\label{eq:bsmumu}
\end{equation}
with
\begin{equation}
C\equiv
\frac{G_F\alpha}{\sqrt{2}\pi}V_{tb}V_{ts}^*\left(\frac{\tan^3\beta}{4 \sin^2\theta_W}\right)
\left[\frac{m_\mu \,m_t\, |\mu|}{M_W^2 (M_{H_1}^2-M_{B_s}^2)}\right]
\left(\frac{\sin 2 \theta_{\tilde{t}}}{2} \right)\Delta f_3
\end{equation}
\noindent where $\Delta f_3=f_3(x_2)-f_3(x_1)$ with
$x_i=m^2_{\tilde{t_i}}/|\mu|^2$ and $f_3(x)=x\log x/(1-x)$ and
$\theta_{\tilde{t}}$ the stop mixing angle~\footnote{Following the
calculation in Ref.~\protect\cite{dedes}, one might get a different
expression than that of Eq. (\protect\ref{eq:bsmumu}), in that the
branching ratio would scale as: $(O_{\phi_1 1}^2+O_{a 1}^2)^2$.} .
In Fig.~\ref{fig:dtl}, we show three contour lines of the rescaled
$\hat{B}(B_s\rightarrow \mu\mu)\equiv (B_s\rightarrow \mu\mu) \times
10^7$: $\hat{B}(B_s\rightarrow \mu\mu)=2$ (solid), 20 (dotted), and
200 (dashed).
For the parameters chosen by combining the results from LEP2 searches,
Thallium EDM, and Bottomonium decay, Eq. (\ref{eq:tanb_mh1}), we get:
$B(B_s\rightarrow \mu\mu)_{CPX}\simeq 6\times 10^{-7}$ taking
$f_{B_s}=0.23$ GeV.  This is three times larger than the present 95 \%
C.L. limit~\cite{bsmumu_limit}: $B(B_s\rightarrow \mu\mu)<2\times
10^{-7}$.  This can be easily made consistent with the present
experimental constraint if some mild cancellation takes place.  The ``GIM
operative point'' mechanism discussed in Ref.~\cite{dedes} may be an
example of such cancellation, when the squark mass matrices are
flavour diagonal. In particular, we find that $B(B_s\rightarrow
\mu\mu)_{CPX}$ is consistent to the experimental upper bound by
choosing $0.8\lsim \rho\lsim 0.9$, where $\rho\equiv
m_{\tilde{q}}/M_{\rm SUSY}$ is the hierarchy factor introduced in
Ref.~\cite{dedes}, with $m_{\tilde{q}}$ the soft mass for squarks of
the first two generations~\footnote{To be consistent with the the
present $B(B_s\rightarrow \mu\mu)$ limit, one may take a point in the
region above the $\hat{B}(B_s\rightarrow \mu\mu)=2$ line in
Fig.~\ref{fig:dtl} where one needs a cancellation in the Thallium EDM.
We note that an appropriate choice of the phases $\Phi_{A_{e,\mu}}$,
$\Phi_{A_{d,s}}$ always allows to make the parameter region of
Eq.~(\protect\ref{eq:tanb_mh1_enlarged}) consistent with the Thallium
EDM constraint when $\rho\simeq$ 0.9. We keep Eq.~(\ref{eq:tanb_mh1})
as our reference point, assuming cancellation in $B(B_s\rightarrow
\mu\mu)$.}.

As far as the SUSY contribution to the anomalous magnetic dipole
moment of the muon $\delta a_{\mu}^{SUSY}$ is concerned, as is well
known uncertainties in the SM make a comparison with the experimental results
difficult. In particular, by combining the SM hadronic vacuum
polarization results obtained from $e^+e^-$ and $\tau^+\tau^-$ data,
the SM calculation turns out to be compatible with observation, and
the following 2--$\sigma$ allowed interval for $\delta a_{\mu}^{SUSY}$
is found~\cite{light_neutralinos}: -160$\lsim \delta
a_{\mu}^{SUSY}\times 10^{11}\lsim 680$. The corresponding contribution
from light neutralinos in the CPX scenario falls comfortably into this
range: $\delta a_{\mu\;CPX}^{SUSY}\times 10^{11}\simeq 1.5$ (to
estimate this we have assumed for the trilinear coupling of the smuon
the same value of the trilinear couplings of the third family given in
Eq. (\ref{eq:mssm-intro-CPXdef})).

\section{The relic density}
\label{section:relicdensity}
Taking into account the latest data from the cosmic microwave data
(CMB) combined with other observations~\cite{wmap3} the 2--$\sigma$
interval for the DM density of the Universe (normalized to the
critical density) is:
\begin{equation}
0.096<\Omega_m h^2<0.122\,,
\label{eq:wmap3}
\end{equation}
\noindent where $h$ is the Hubble parameter expressed in units of 100
km s$^{-1}$ Mpc$^{-1}$. In Eq.(\ref{eq:wmap3}) the upper bound on
$\Omega_m h^2$ establishes a strict upper limit for the abundance of
any relic particle.  In particular, the neutralino relic abundance is
given by the usual expression:
\begin{equation}
\Omega_{\chi} h^2 = \frac{x_f}{{g_{\star}(x_f)}^{1/2}} \frac{3.3 \cdot
10^{-38} \; {\rm cm}^2}{\widetilde{<\sigma_{ann} v>}},
\label{eq:omega}
\end{equation}
\noindent where $\widetilde{<\sigma_{ann} v>} \equiv x_f {\langle
\sigma_{\rm ann} \; v\rangle_{\rm int}}$, ${\langle \sigma_{\rm ann}
\; v\rangle_{\rm int}}\equiv\int_{T_0}^{T_f} \langle \sigma_{\rm ann}
\; v\rangle \;dT/m_{\chi}$ being the integral from the present
temperature $T_0$ up to the freeze-out temperature $T_f$ of the
thermally averaged product of the annihilation cross-section times the
relative velocity of a pair of neutralinos $\sigma_{\rm ann} \; v$,
$x_f$ is defined as $x_f \equiv \frac{m_{\chi}}{T_f}$ and
${g_{\star}(x_f)}$ denotes the relativistic degrees of freedom of the
thermodynamic bath at $x_f$. For the determination of $x_f$ we adopt a
standard procedure ~\cite{kolb_turner}.

In absence of some resonant effect, the natural scale of the
annihilation cross section times velocity $\sigma_{ann} v$ of CPX
light neutralinos is far too small to keep the relic abundance below
the upper bound of Eq.(\ref{eq:wmap3}) (in particular they are very
pure $B$--inos and their mass is below the threshold for annihilation
to bottom quarks, which is usually the dominant channel of
$\sigma_{ann} v$ for light neutralinos
~\cite{light_neutralinos}). However, when $m_{\chi}\simeq M_{H_1}/2$
neutralinos annihilate through the resonant channel
$\chi\chi\rightarrow H_1\rightarrow standard\; particles$, bringing
the relic abundance down to acceptable values. In the Boltzmann
approximation the thermal average of the resonant $\sigma_{ann} v$ to
the final state $f$ can be obtained in a straightforward way from the
following relation among interaction rates:
\begin{equation}
\frac{n_{\chi}^2}{2}<\sigma_{ann} v>_{{\rm res},f} = <\Gamma(\chi\chi\rightarrow f)>=<\Gamma(\chi\chi\rightarrow
  H_1)B(H_1\rightarrow f)>=n_{H_1}\Gamma_{\chi}\frac{K_1(x_{H_1})}{K_2(x_{H_1})}B_f,
\label{eq:resonant}
\end{equation}
\noindent where brackets indicate thermal average, $\Gamma_{\chi}$ is
the zero--temperature $H_1$ annihilation amplitude to neutralinos and
the thermal average of this quantity is accounted for by the ratio of
modified Bessel functions of the first kind $K_1$ and $K_2$, $B_f$ is
the $H_1$ branching ratio to final state $f$, $n_i=g_i m_i^3
K_2(x_i)/(2 \pi^2 x_i)$ are the equilibrium densities with
$x_i=m_i/T$, $T$ the temperature, and $g_i$ the corresponding internal
degrees of freedom, $g_{\chi}=2$, $g_{H_1}$=1. The factor of 1/2 in
front of Eq.(\ref{eq:resonant}) accounts for the identical initial
states in the annihilation.  From Eq.(\ref{eq:resonant}), and summing
over final states $f$, one gets:
\begin{equation}
<\sigma_{ann} v>_{\rm res}=\frac{\pi^2 M_{H_1}^2}{
  m_{\chi}^5}\frac{x_{\chi}
  K_1(x_{H_1})}{K_2^2(x_{\chi})}\Gamma(H_1)B_{\chi}
  (1-B_{\chi})\Theta\left (\frac{x_{H_1}}{x_{\chi}}-2 \right),
\label{eq:sigmav}
\end{equation}
\noindent with $B_{\chi}=\Gamma_{\chi}/\Gamma(H_1)$ , $\Gamma(H_1)$
the total decay amplitude of $H_1$, while $\Theta$ is the Heaveside
step function~\footnote{Our result for $<\sigma_{ann} v>_{\rm res}$ is
a factor of 2 larger than the corresponding expression calculated in
~\cite{Gondolo}. We explain this discrepancy with the fact that in
~\cite{Gondolo} the Breit--Wigner expression of the cross section,
Eq.(6.1), should be a factor 2 larger, to compensate for a factor 1/2
contained in the decay amplitude of the resonance to 2 (identical)
neutralinos}.

By making use in Eq.(\ref{eq:sigmav}) of the approximations $K_1(z)\simeq K_2(z)\simeq
(\pi/(2z))^{1/2}\exp(-z)$, valid for $z\gg 1$, the integral over
temperature can be done analytically, leading to:

\begin{equation}
\widetilde{<\sigma_{ann} v>}_{\rm res}\simeq 4\pi^2\frac{x_f
\Gamma(H_1)}{m_{\chi}^3}\frac{B_{\chi}(1-B_{\chi})}{\beta_{\chi}}
\sqrt{\frac{\delta(\delta+1)}{2}}\left[1-erf\left(
  \sqrt{2(\delta-1)x_f}\right) \right ],
\label{eq:sigmatilde}
\end{equation}
\noindent where $\delta\equiv M_{H_1}/(2 m_{\chi})$,
$\beta_{\chi}=\sqrt{1-\delta^{-2}}$ and $erf(x)=2/\sqrt{\pi}\int_0^x
exp(-t^2)\;dt$.

\begin{figure}[t]
\vspace{0.0cm}
\centerline{\epsfig{figure=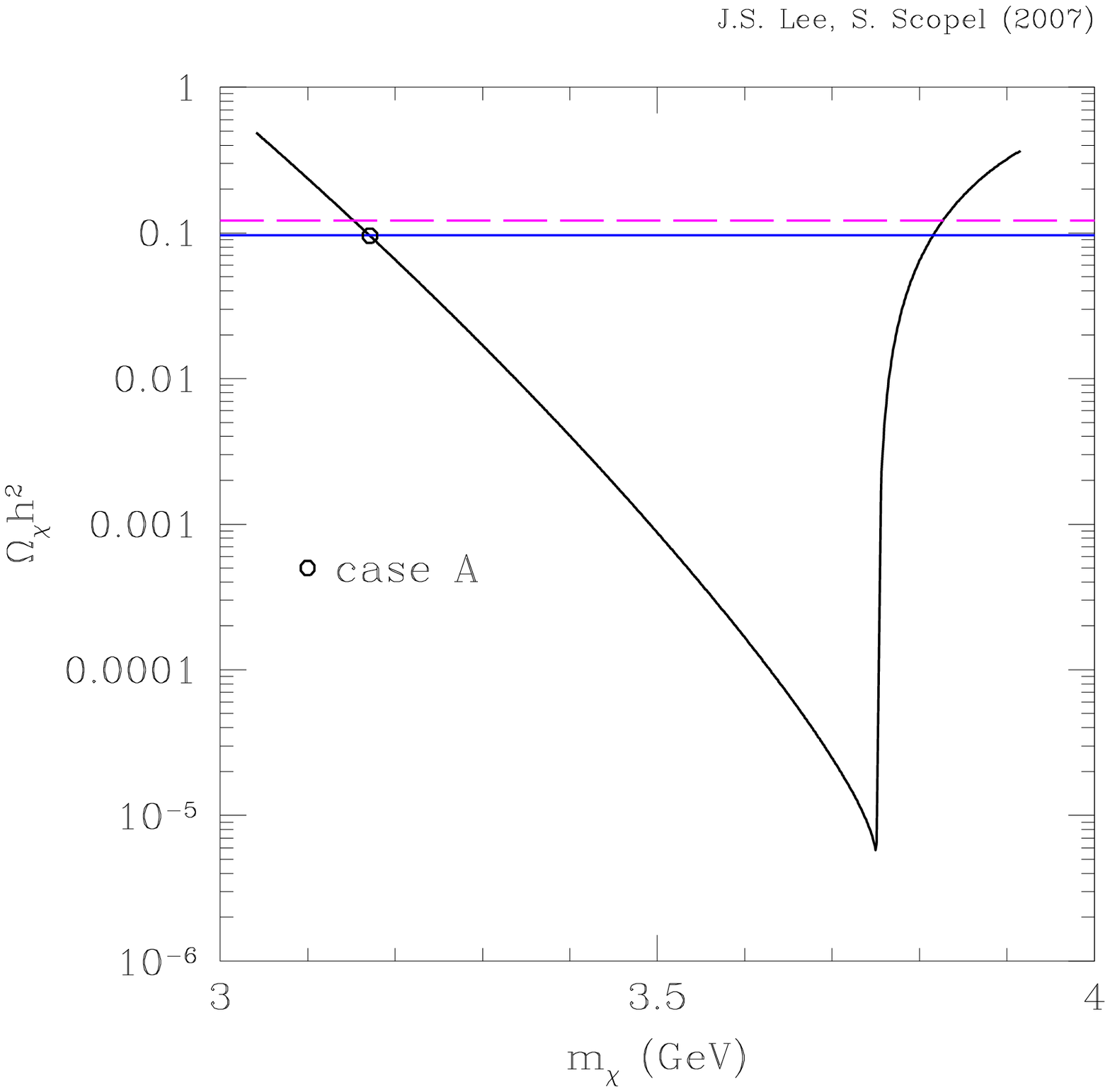,width=8cm,bbllx=17, bblly=175,
    bburx=504,  bbury=666}}
\vspace{-0.5cm}
\caption{{\it Relic abundance as a function of the neutralino mass
$m_{\chi}$ for the CPX scenario with $M_{H_1}$=7.5 GeV, $\tan\beta$=3,
$M_{\rm SUSY}$=0.5 TeV and $\Phi_{A}=\Phi_3=90^\circ$.  The two
horizontal lines indicate the interval of Eq.(\protect\ref{eq:wmap3}).
The circle, where $\Omega_{\chi} h^2=(\Omega_m h^2)_{min}$, is
discussed as ``case A'' in Section \protect\ref{section:dmsearches}.
\label{fig:mchi_omega}}}
\end{figure}

The result of our calculation is shown in Fig.~\ref{fig:mchi_omega},
where the neutralino relic abundance $\Omega_{\chi} h^2$ is shown as a
function of the mass $m_{\chi}$. In this calculation the annihilation
cross section has been calculated including the off--resonance
contribution to the annihilation cross section. In this way we have
checked that, for $m_{\chi}<M_{H_1}/2$, $\widetilde{<\sigma_{ann}
v>}_{\rm res}$ is always the dominant contribution in the calculation
of $\Omega_{\chi} h^2$, and Eq.(\ref{eq:sigmatilde}) is an excellent
approximation of $\widetilde{<\sigma_{ann} v>}$.  The off--resonance
contribution to the annihilation cross section is also responsible for
regularizing the relic density for $m_{\chi}>M_{H_1}/2$, where
$\widetilde{<\sigma_{ann} v>}_{\rm res}$ is vanishing. The asymmetric
shape of the curve in Fig.~\ref{fig:mchi_omega} is due to the fact that
thermal motion allows neutralinos with $m_{\chi}<M_{H_1}/2$ to reach
the center--of--mass energy needed to create the resonance, while this
is not possible for $m_{\chi}>M_{H_1}/2$. In the same figure, the two
horizontal lines indicate the range of Eq.(\ref{eq:wmap3}). The point
shown as a circle and indicated as ``case A'' will be used in the
following section as a representative point to calculate some signals.

In Fig.~\ref{fig:mchi_omega} the neutralino mass range allowed by
cosmology is: 3.15 GeV $\simeq m_{\chi} \simeq$ 3.83 GeV. Allowing for
the variation of $M_{H_1}$ within the range of
Eqs.(\ref{eq:tanb_mh1},\ref{eq:tanb_mh1_enlarged}), this range is
enlarged to:
\begin{equation}
2.93 \;\;{\rm GeV} \lsim m_{\chi}\lsim 5\;\; {\rm GeV}.
\label{eq:mass_range}
\end{equation}
\noindent In this scenario the neutralino relic abundance can fall in
the range of Eq.(\ref{eq:wmap3}) only with some level of tuning at the
boundaries of the allowed mass range. For intermediate values of
$m_{\chi}$ either the neutralino is a sub--dominant component of the
DM, or some non--thermal mechanism for its cosmological density needs
to be introduced.  Of course all our considerations are valid if
standard assumptions are made about the evolution of the early
Universe (e.g. about the reheating temperature at the end of
inflation, the energy budget driving Hubble expansion, entropy
production, etc).

\section{Dark matter searches}

\label{section:dmsearches}
Neutralinos are CDM particles, and are supposed to clusterize at the
galactic level. This implies that they can provide the DM density
which is gravitationally measured in our Galaxy. In particular, in
this section we will assume for the DM density in the neighborhood of
the solar system the reference value $\rho_0$= 0.3 GeV/cm$^3$.  When
we compare our calculation of $\Omega_{\chi}h^2$ to the interval of
Eq.(\ref{eq:wmap3}), we interpret the lower bound on the DM relic
density $(\Omega_m h^2)_{min}$=0.096 as the average abundance below
which the halo density of a specific CDM constituent has to be
rescaled as compared to the total CDM halo density. So, whenever
$\Omega_{\chi}h^2<(\Omega_m h^2)_{min}$ we assume that neutralinos
provide a local density $\rho_{\chi}$ which is only a fraction of
$\rho_0$.  For the determination of the rescaling factor
$\xi\equiv\rho_{\chi}/\rho_0$ we adopt the standard recipe:

\begin{equation}
\xi=min[1,\Omega_{\chi} h^2/(\Omega_m h^2)_{min}].
\label{eq:rescaling}
\end{equation}

Neutralinos in the halo of our Galaxy can be searched for through
direct and indirect methods. In particular, CPX light neutralinos are
quite hard to detect through direct detection.  Direct detection
consists in the measurement of the elastic scattering of neutralinos
off the nuclei of a low--background detector. For the mass range of
Eq.(\ref{eq:mass_range}), the most stringent upper bound on the
neutralino--nucleon coherent elastic cross section $\sigma_{\rm
scalar}^{\rm (nucleon)}$ is provided by the CRESST-1 experiment
~\cite{CRESST}, $\sigma_{\rm scalar}^{\rm (nucleon)}\lsim $10$^{-38}$
cm$^2$~\footnote{We remind that the DAMA Collaboration measures an
annual modulation effect \protect\cite{dama}, compatible to what
expected by relic neutralinos in some supersymmetric models
\protect\cite{susy_modulation} or by other dark matter candidates
\protect\cite{dama2}. The model discussed in the present paper
provides elastic cross-sections too low to explain the DAMA modulation
effect.}. This value is much above the cross section expected in our
scenario, which falls in the range $(\sigma_{\rm scalar}^{\rm
(nucleon)})_{CPX}\simeq$ 10$^{-42}$ cm$^2$. This is due to the fact
that the neutralino-Higgs coupling which dominates this process is
suppressed, since $\tan\beta$ is small and $\chi$'s are very pure
$B$-inos due to the large value of $|\mu|$. Moreover, $\sigma_{\rm
scalar}^{\rm (nucleon)}$ is dominated by the exchange of scalar Higgs
bosons, while $H_1$ is mostly pseudoscalar. Finally, contrary to
annihilation, no resonant enhancement is present in the elastic cross
section, since scattering proceeds through $t$--channel.

For this reason in this section we concentrate on the indirect
detection of CPX light neutralinos. In our scenario the neutralino
relic density $\Omega_{\chi} h^2$ is driven below the observational
limit by the resonant enhancement of the annihilation cross section
$\widetilde{<\sigma_{ann} v>}$.  The same cross section calculated at
present times, $<\sigma_{ann} v>_0$, enters into the calculation of
the annihilation rate of neutralinos in our galaxy. This could produce
observable signals, like $\gamma$'s, $\nu$'s or exotic components in
Cosmic Rays (CR), like antiprotons, positrons, antideuterons. 

Note, however, that one can have $<\sigma_{ann}
v>_0\ll\widetilde{<\sigma_{ann} v>}$. In fact, as already shown in
Section \ref{section:relicdensity}, the thermal motion in the early
Universe ($x_{\chi}\simeq x_f\simeq 20$) allows neutralino resonant
annihilation when $m_{\chi}<M_{H_1}/2$. However, for the same
neutralinos the contribution of the resonance to $<\sigma_{ann} v>_0$
can be negligible at present times, since their temperature in the
halo of our Galaxy is of order $x_{\chi,0}\simeq$10$^{-6}\ll
x_f$. This implies that the annihilation cross section can be large
enough in the early Universe in order to provide the correct relic
abundance, but not so large at present times as to drive indirect
signals beyond observational limits.

In the following we will discuss expected signals for $\gamma$ rays
and antiprotons. The results of our analysis are summarized in
Figs.~\ref{fig:gamma_line},\ref{fig:mchi_gamma_egret},\ref{fig:pbar}
and \ref{fig:glast}.  In all figures observables are calculated in the
CPX scenario with $M_{H_1}$=7.5 GeV, $\tan\beta$=3, $M_{\rm SUSY}$=0.5
TeV and $\Phi_{A}=\Phi_3=90^\circ$, and plotted as a function of the
neutralino mass $m_{\chi}$. The solid lines show our results obtained
by adopting the rescaling procedure for the local density explained
above, while for comparison, dashed lines are calculated assuming
$\rho_{\chi}=\rho_0$. When rescaling is applied, indirect signals are
proportional to the combination $\xi\rho_{\chi}^2<\sigma_{ann} v>_0$,
so they reach their maximum value when $\Omega_{\chi} h^2=(\Omega_m
h^2)_{min}$.

For all our results we have adopted, as a reference model, a
Navarro-Frenk-White (NFW) profile for the DM density:

\begin{equation}
\rho(r)=\rho_0 \frac{(r_0/a)^{\gamma}[1+(r_0/a)^{\alpha}]^{(\beta-\gamma)/\alpha}}
{(r/a)^{\gamma}[1+(r/a)^{\alpha}]^{(\beta-\gamma)/\alpha}}\,,
\label{eq:nfw}
\end{equation}

\noindent where $r_0=8.5$ kpc is the distance of the solar system from
the Galactic Center (GC), $a=25$ kpc is the core radius, while
$(\alpha,\beta,\gamma)=(1,3,1)$.

\subsection{Gamma flux from the Galactic Center}

\begin{figure}[t]
\vspace{0.0cm}
\centerline{\epsfig{figure=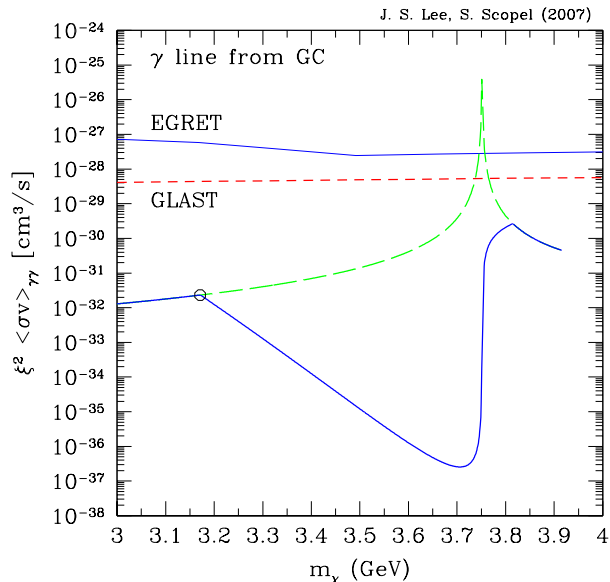,width=8cm,bbllx=27, bblly=175,
    bburx=504,  bbury=666}}
\vspace{-0.5cm}
\caption{{\it Rescaled, zero--temperature neutralino annihilation
cross section to two photons, as a function of $m_{\chi}$. For the
solid line the rescaling factor $\xi$ is calculated according to
Eq.~(\protect\ref{eq:rescaling}), while for the dashed one
$\xi=1$. The solid horizontal line shows the corresponding constraint
from a search for a $\gamma$ line from the GC from EGRET
\protect\cite{EGRET_gammaline}. The dashed horizontal line is an
estimate for the sensitivity of GLAST \protect\cite{glast} for a
similar search, when model $A-N2$ of Ref.\protect\cite{gc_background}
is used to extrapolate HESS data to lower energies.  Circle: see
caption of Fig.~\protect\ref{fig:mchi_omega}.
\label{fig:gamma_line}}}
\end{figure}

\begin{figure}[t]
\vspace{0.0cm}
\centerline{\epsfig{figure=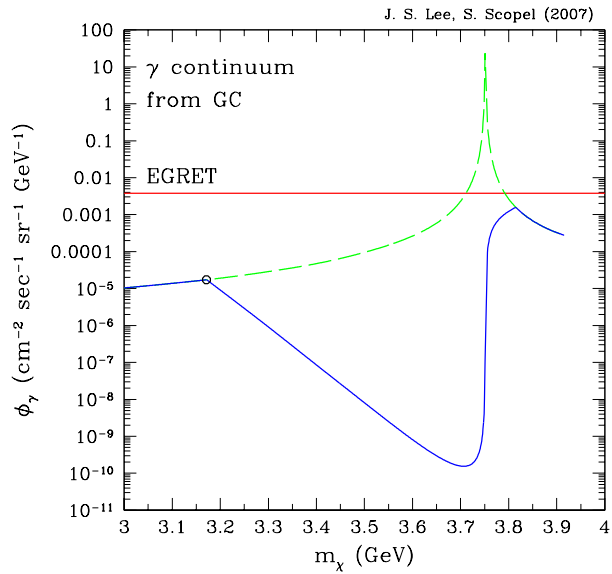,width=8cm,bbllx=22, bblly=175,
    bburx=504,  bbury=666}}
\vspace{-0.5cm}
\caption{{\it Continuum $\gamma$ flux from the GC calculated for
$E_{\gamma}$=122 MeV. For the solid line the neutralino local density
is rescaled according to Eq.~(\protect\ref{eq:rescaling}), while for
the dashed one, $\xi=1$. The fluxes are compared to the corresponding
measurement from EGRET \protect\cite{EGRET}, shown as the horizontal
solid line.  Circle: see caption of Fig.~\protect\ref{fig:mchi_omega}.
\label{fig:mchi_gamma_egret}}}
\end{figure}

In Figs.~\ref{fig:gamma_line} and \ref{fig:mchi_gamma_egret} we show our
results for a $\gamma$ signal from the GC, which is the most promising
source of $\gamma$' from neutralino annihilation.  Note that signals
from the GC are proportional to the line--of--sight integral (i.e.,
performed on a path pointing from the observer to the $\gamma$ source)
$\bar{J}\equiv 1/{\Delta\Omega}\int_{l.o.s.}  \rho_{\chi}^2 \;dl
\;d\Omega$, where $\Omega$ is the pointing angle of observation in the
sky and $\Delta\Omega$ is the experimental angular resolution. The
quantity $\bar{J}$ is very sensitive to the particular choice of
density profile, and may span several orders of magnitude, especially
for those models that diverge in the origin, as for the NFW, where a
cut--off radius $r_{cut}$ is needed ~\cite{pieri}. In particular, in
our calculation we use: $r_{cut}=10^{-2}$ pc.

The $\gamma$ signal from neutralinos takes two contributions: a line
with $E_{\gamma}=m_{\chi}$, produced by direct annihilation of
$\chi$'s to two $\gamma$'s, and a continuum, which is mainly due to
the annihilation of $\pi^0$'s produced in the fragmentation and decay
of other final states (quarks,gluons and $\tau$'s).

In the first case, the zero--temperature annihilation cross section
of neutralinos to photons $<\sigma_{ann} v>_{0,\gamma\gamma}$ is
usually suppressed, since it takes place at the one--loop
level. However, in our case the contribution of the $H_1$ resonance,
$\chi\chi\rightarrow H_1\rightarrow\gamma\gamma$ can lead to a strong
enhancement of the signal, as is evident from
Fig.~\ref{fig:gamma_line}, where our calculation of $\xi^2<\sigma_{ann}
v>_{0,\gamma\gamma}$ is compared to the upper bound (horizontal solid
line) on the same quantity from EGRET~\cite{EGRET_gammaline} (such
analysis has been performed on a region of
10$^{\circ}\times$10$^{\circ}$ around the GC, which, for our
assumptions, implies $\bar{J}\simeq$ 120 GeV$^2$ cm$^{-6}$ kpc).  In
this figure the solid line shows our result when $\xi^2$ is calculated
according to Eq.(\ref{eq:rescaling}), while for the dashed line
$\xi=1$.

In the same figure we also show with a horizontal dashed line an
estimate for the prospect of detection of the same quantity with GLAST
(in this case we have assumed an angular resolution of
$\Delta=10^{-5}$ sr in the calculation of $\bar{J}$, leading to
$\bar{J}\simeq$ 2500). This estimate is somewhat uncertain, since HESS
~\cite{hess} has detected a TeV source of $\gamma$'s in the GC which
is likely to be of standard origin, representing a background for DM
searches, and potentially making detection of new physics in that
region more difficult~\cite{gc_background}.  Our estimate of the
background is obtained by extrapolating the HESS source to lower
energies by using the model $A-N2$ described in ~\cite{gc_background}
(we have made the same assumptions also to estimate the horizontal
solid line in Fig.~\ref{fig:glast}). Our choice of model $A-N2$ is
optimistic, since it implies the smallest extrapolated background at
low energies among those discussed in ~\cite{gc_background}. Of
course, a more conservative choice for the model adopted to explain
the HESS source could make prospects of DM detection for GLAST in the
GC much worse.

As far as the continuum signal is concerned, we have calculated the
$\gamma$ yield from the final states of neutralino annihilation using
{\tt PYTHIA}~\cite{pythia}. The result of the calculation is shown in
Fig.~\ref{fig:mchi_gamma_egret}, where we have evaluated the $\gamma$
flux $\phi_{\gamma}$ from the GC for $E_{\gamma}=122$ MeV, and
compared it with the corresponding flux measurement from EGRET
~\cite{EGRET}, shown as a horizontal solid line (this particular
energy bin is within the range where the data are well fitted by a
standard background. In this case the flux has been measured in an
angular region of 10$^{\circ}\times$4$^{\circ}$ around the GC, which
corresponds, for our choice of parameters, to $\bar{J}\simeq$ 184
GeV$^2$ cm$^{-6}$ kpc).

From Figs.~\ref{fig:gamma_line},\ref{fig:mchi_gamma_egret} we can
conclude that, with reasonable choices for the DM density profile,
$\gamma$ signals in our scenario are compatible with observations.  In
both figures we have indicated with a circle the point indicated as
``case A'' in Fig.~\ref{fig:mchi_omega}.

\subsection{Antiproton flux}
\label{section:pbars}

\begin{figure}[t]
\vspace{0.0cm}
\centerline{
\epsfig{figure=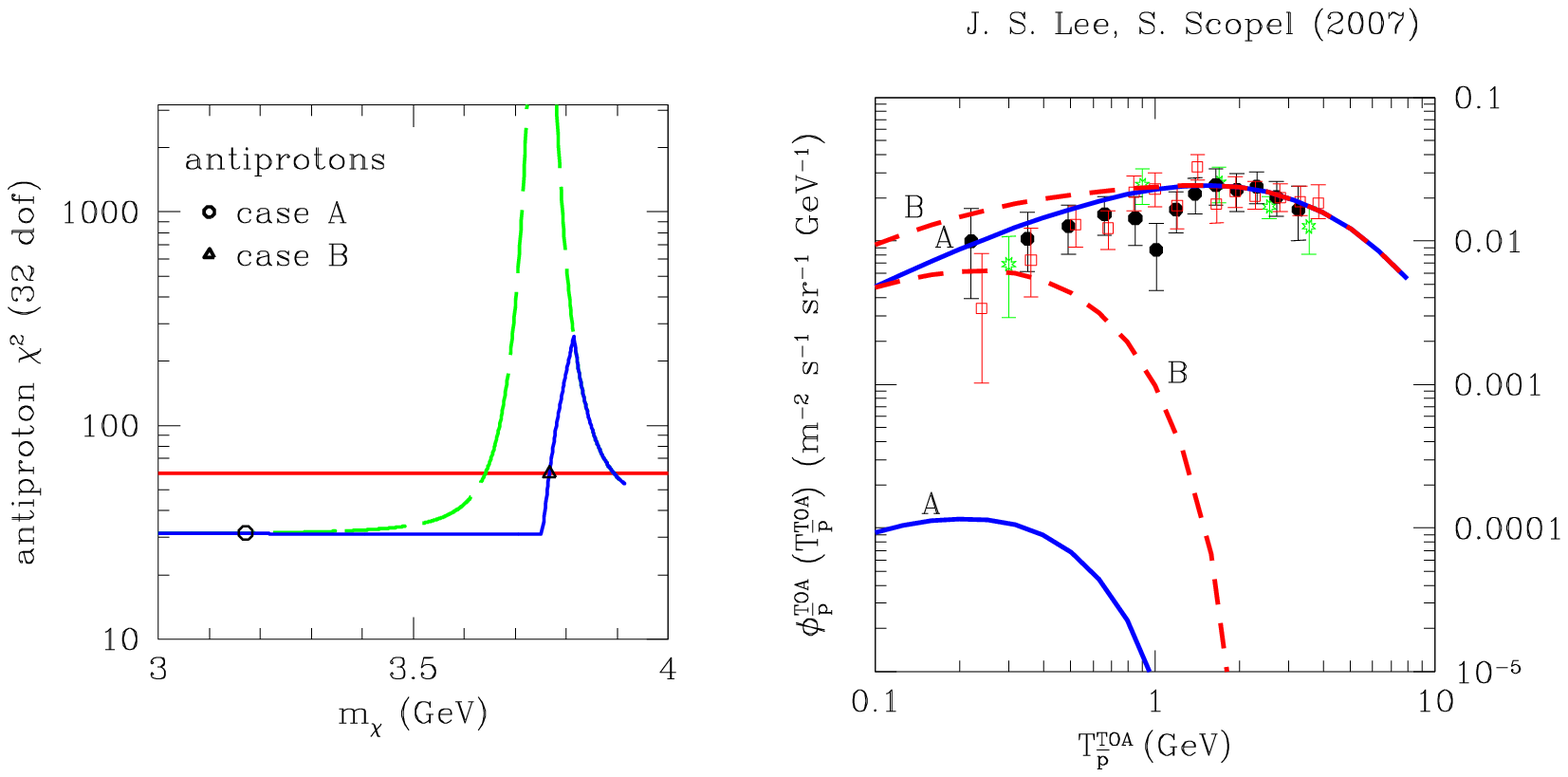,width=14cm,bbllx=64, bblly=193,
    bburx=544,  bbury=436}
}
\vspace{-0.5cm}
\caption{{\it {\bf Left:} $\chi^2$ calculated for the
top--of--atmosphere $\bar{p}$ flux compared to the experimental data
from BESS and AMS \protect\cite{pbar_data} (explicitly shown in the
right--hand panel). For the solid line the neutralino local density is
rescaled as explained in Section \protect\ref{section:dmsearches}. For
the dashed one, $\xi=1$.  The solid horizontal line indicates the 99.5
\% C.L. upper bound for the $\chi^2$. The comparison of the $\chi^2$
with its upper limit allows to set an upper bound to $m_{\chi}$,
indicated by case B and shown as a triangle.  On the other hand the
circle indicates case A, introduced in
Fig.~\protect\ref{fig:mchi_omega}.{\bf Right:} Top--of--atmosphere
antiproton flux as a function of the kinetic TOA energy of
$\bar{p}$'s, for the cases $A$ (solid curve) and $B$ (dashes), shown
in the left panel as a circle and a triangle, respectively. In both
cases the lower curve shows the contribution from primaries produced
by neutralino annihilation, while the upper curve is the total flux
where the primary contribution is added to the standard secondary
one. The 32 experimental points \protect\cite{pbar_data} are the same
that are used to calculate the $\chi^2$ shown in the left--hand panel:
full circles: BESS 1995-97; open squares: BESS 1998; stars: AMS.
\label{fig:pbar}}}
\end{figure}

As far as light neutralinos are concerned, a particularly stringent
limit is provided by the flux of primary antiprotons that are produced
from the hadronization of neutralino--annihilation final states
~\cite{indirect,pbars,propagation}. This is due to the fact that the
$\bar{p}$ flux observed experimentally is quite in agreement with that
expected from $\bar{p}$ secondary production from cosmic
rays~\cite{pbar_data}, so that not much room is left for exotic
contributions. Moreover, as all other annihilation processes, the
primary neutralino signal scales with the neutralino number density
$\propto 1/m_{\chi}^2$, so it is enhanced for light masses. However,
once they are produced in the DM halo, primary $\bar{p}$'s interact
with the magnetic field of the Galaxy, and a complex propagation model
is needed in order to calculate the fraction of them that reaches the
Earth. Unfortunately, the main parameters of the propagation model are
fixed by using secondary CR data (such as the $B$/$C$ ratio) which
mainly depend on the galactic disk, while primary $\bar{p}$'s from
neutralino annihilation are produced in the galactic halo. This
induces uncertainties in the primary flux as large as two orders of
magnitude ~\cite{pbars}.  In particular, for our analysis we have used
the public code provided by Ref.~\cite{propagation} for the $\bar{p}$
propagation. Although the code of Ref.~\cite{propagation} is a
simplified one, where in particular energy--redistribution effects are
neglected, it serves well our needs for checking the viability of our
scenario. In the left panel of Fig.~\ref{fig:pbar} we show the
$\chi^2$ calculated by comparing the sum of the primary and secondary
top-of-atmosphere (TOA) $\bar{p}$ fluxes to the experimental data from
BESS and AMS. In particular, for the calculation of the $\chi^2$ we
have used the same 32 data points from BESS 1995-97, BESS 1998 and AMS
1998~\cite{pbar_data} shown in the right panel. For our calculation we
have used a solar modulation parameter $\phi$= 500 MV, corresponding to
the period of minimal solar activity when these experiments have taken
data. Moreover, in order to be conservative, we have used the minimal
propagation model in Table III of Ref.~\cite{propagation},
i.e. $\delta$=0.85, $K_0$=0.0016 kpc$^2$ Myr$^{-1}$, $L$=1 kpc,
$V_c$=13.5 km/s and $V_A=$ 22.4 km/s for the propagation parameters
$\delta$ and $K_0$, for the size of the diffusion zone $L$, and for
the galactic wind $V_c$.  In absence of a SUSY contribution we find
$\chi^2\simeq$30, which confirms the good agreement between the data
and the standard secondary production (for this latter flux we use the
quantity calculated in ~\cite{propagation}). As a conservative upper
bound for the $\chi^2$ we take $\chi^2=60$, which for 32 degrees of
freedom implies a statistical disagreement at the level of 99.5
\%. This value is shown in Fig.~\ref{fig:pbar} as a solid horizontal
line.

\begin{figure}[t]
\vspace{0.0cm}
\centerline{\epsfig{figure=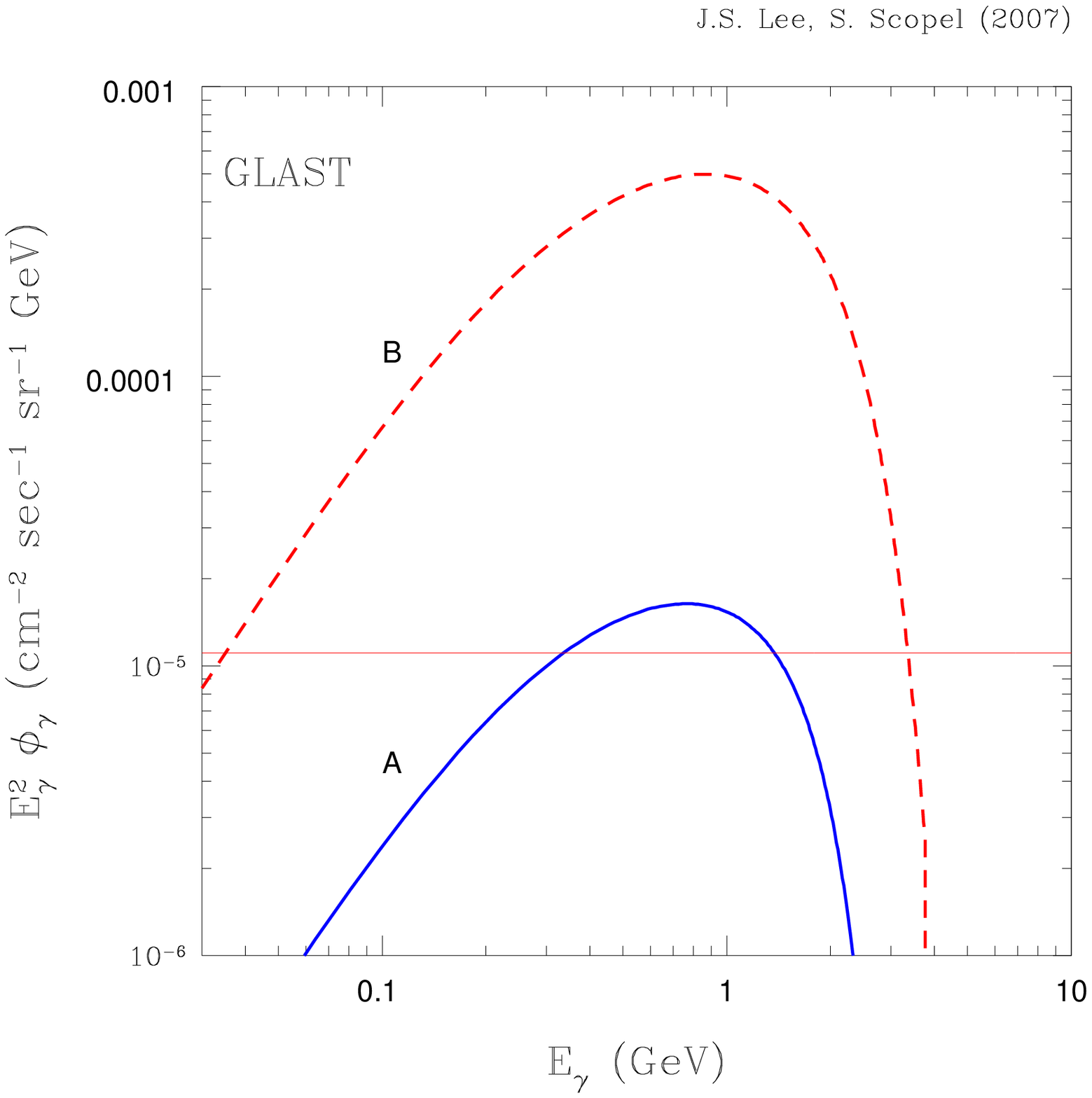,width=8cm,bbllx=22, bblly=175,
    bburx=504,  bbury=666}}
\vspace{-0.5cm}
\caption{{\it Expected signal for GLAST \protect\cite{glast} for a
gamma continuum flux from the Galactic Center as a function of the
$\gamma$ energy for the cases $A$ (solid curve) and $B$ (dashes)
indicated in Fig.~\protect\ref{fig:mchi_omega},
\protect\ref{fig:gamma_line} and \protect\ref{fig:mchi_gamma_egret} as
a circle and a triangle, respectively. The horizontal line is an estimate
of the background for GLAST (see text).
\label{fig:glast}}}
\end{figure}

From Fig.~\ref{fig:pbar} one can see that, even when the rescaled
neutralino local density is used, the $\chi^2$ from $\bar{p}$ data
exceeds 60 for $m_{\chi}\simeq$ 3.77 GeV. This values for the
neutralino mass is indicated with a triangle (case B), along with the
circle which indicates case A introduced in Fig.~\ref{fig:mchi_omega}.
Both cases are shown in the right--hand panel, where the upper curves
show the total $\bar{p}$ fluxes, while the lower curves show the SUSY
contributions.


From the discussion of this section, where $m_{H_1}$=7.5 GeV and
$\tan\beta$=3, we obtain for $m_{\chi}$ the allowed range: 3.15 GeV
$\lsim m_{\chi} \lsim$ 3.77 GeV. The boundaries of this range
correspond to the cases A and B introduced previously. Assuming for
$M_{H_1}$ and $\tan\beta$ the range of Eqs. (\ref{eq:tanb_mh1},
\ref{eq:tanb_mh1_enlarged}) this interval for $m_{\chi}$ is enlarged
to 2.93 GeV $\lsim m_{\chi} \lsim$ 5 GeV.

We conclude this section by showing and example for the prospects of
detection for cases A and B in future DM searches: in
Fig.~\ref{fig:glast} the $\gamma$ continuum flux from the GC is
estimated for GLAST (we have assumed here an angular resolution of
$\Delta=10^{-5}$ sr in the calculation of $\bar{J}$, leading to
$\bar{J}\simeq$ 2500).  This flux is compared to an estimate of the
background for the same detector, shown as a horizontal line,
calculated with the same assumptions as in Fig.~\ref{fig:gamma_line},
where model $A-N2$ of Ref.\protect\cite{gc_background} is used to
extrapolate HESS data to lower energies.

\section{Conclusions}

In the present paper we have discussed the lower bound to the lightest
Higgs boson $H_1$ in the MSSM with explicit CP violation, and the
phenomenology of the lightest relic neutralino in the same scenario.
In particular, we have examined the parameter space region
$M_{H_1}\lsim 10$ GeV and $3\lsim \tan\beta \lsim 10$, 
in the CPX scenario, an interval which has not been excluded by
the combined searches of the four LEP collaborations.
%
We find that the combination of experimental constraints coming from
Thallium EDM measurements and quorkonium decays restricts the region
allowed by LEP to: $7~{\rm GeV} \lsim M_{H_1}\lsim 7.5$ GeV, $\tan\beta
\simeq 3$. In this range, the branching ratio $B(B_s\rightarrow
\mu\mu)$ is compatible to the present experimental upper bounds
provided some moderate cancellation is allowed between the stop--loop
contribution and that of other squarks. Furthermore, the allowed
parameter space can be relaxed to $7~{\rm GeV}\lsim M_{H_1} \lsim
10~{\rm GeV}$ and $3\lsim \tan\beta \lsim 5$ if a cancellation less
severe than 1 part in 10 is also assumed in Thallium EDM between
two--loop contributions and, for example, those depending from first-- and
second--generation phases.

For the above choice of parameters and assuming a departure from the
usual GUT relation among gaugino masses ($|M_1| \ll |M_2|$) we find
that neutralinos with $2.9~{\rm GeV} \lsim m_{\chi}\lsim 5~{\rm GeV}$
can be viable DM candidates. We refer to them as CPX light
neutralinos.  In particular, in the CPX scenario the neutralino is a
very pure $B$-ino configuration, suppressing its Higgs--mediated cross
sections. However, through resonant annihilation to $H_1$ the thermal
relic density of neutralinos with $m_{\chi}\simeq M_{H_1}/2$ can be
either tuned within the range compatible to WMAP or virtually erased,
allowing for alternative non-thermal mechanisms.  The cosmologically
allowed range for $m_{\chi}$ extends to $m{\chi}\lsim M_{H_1}/2$ due
to the effect of thermal motion in the early Universe.
%
We also have discussed the phenomenology of CPX light neutralinos,
showing that signals for indirect Dark Matter searches
are compatible with the present experimental constraints, as long as
$m_\chi\lsim M_{H_1}/2$. On the other hand, part of the range
$m_\chi\gsim M_{H_1}/2$ allowed by cosmology is excluded by antiproton
fluxes.


Finally, we note that our study shares some generic features with other
models such as the next--to--minimal supersymmetric model (NMSSM), in
which a light Higgs boson may escape LEP searches and the observed amount
of DM is explained through neutralino resonant annihilation
~\cite{NMSSM1,NMSSM2}.


\section*{Acknowledgments}
The work of J.S.L. was supported in part by 
the Korea Research Foundation (KRF)
and the Korean Federation of Science and Technology Societies Grant
and in part by the KRF grant KRF--2005--084--C00001
funded by the Korea Government (MOEHRD, Basic Research Promotion Fund).
S.S. would like to thank A. Bottino for useful discussions.


\begin{thebibliography}{99}
\bibitem{glast} See for instance, 
\verb+ www-glast.slac.stanford.edu/software/IS/glast_lat_performance.htm+.
\end{thebibliography}
\end{document}